\documentclass[12pt]{article}
\usepackage{graphicx}
\usepackage{authblk}
\usepackage[hyphens]{url}   

\usepackage[sorting=none]{biblatex}
\usepackage{setspace}

\usepackage{listings}
\lstdefinelanguage{D}
    {morekeywords={auto},
    sensitive=false,
    morecomment=[l]{///},
    morecomment=[l]{//},
    morecomment=[s]{/*}{*/},
    morestring=[b]",
}

\usepackage{hyperref}
\hypersetup{
    colorlinks=true,
    urlcolor=blue,
    citecolor=magenta,
}

\title{Typesafe Coordinate Systems in High-Throughput Sequencing Applications}

\author[1]{\small Charles Thomas Gregory}
\author[1,2,3]{\small James S. Blachly}

\affil[1]{\footnotesize The Ohio State University Comprehensive Cancer Center}
\affil[2]{\footnotesize The Ohio State University Department of Internal Medicine, Division of Hematology}
\affil[3]{\footnotesize The Ohio State University Department of Biomedical Informatics}

\begin{document}

\maketitle

\begin{abstract}
High-throughput sequencing file formats and tools encode coordinate intervals with respect to a reference sequence in at least four distinct, incompatible ways. Integrating data from and moving data between different formats has the potential to introduce subtle off-by-one errors. Here, we introduce the notion of typesafe coordinates: coordinate intervals are not only an integer pair, but members of a type class comprising four types: the Cartesian product of a zero or one basis, and an open or closed interval end. By leveraging the type system of statically and strongly-typed, compiled languages we can provide static guarantees that an entire class of error is eliminated. We provide a reference implementation in D as part of a larger work (dhtslib), and proofs of concept in Rust, OCaml, and Python. Exploratory implementations are available at \url{https://github.com/blachlylab/typesafe-coordinates}.
\end{abstract}

\doublespacing

\section*{Background}

Bioinformaticians working in genomics employ a broad variety of file formats to aid in definition, organization, aggregation, and analysis of sequencing data. One universal aspect of myriad formats is the use of coordinates and intervals based on a reference sequence, typically a genome. Troublingly, different standards do not represent coordinates’ relationship to the underlying reference sequence uniformly. This discordance among formats which are otherwise frequently used together in workflows is a common stumbling block for junior practitioners, and off-by-one errors may affect even experienced bioinformaticians and computational biologists.

\begin{figure}[h]
\includegraphics[width=\columnwidth]{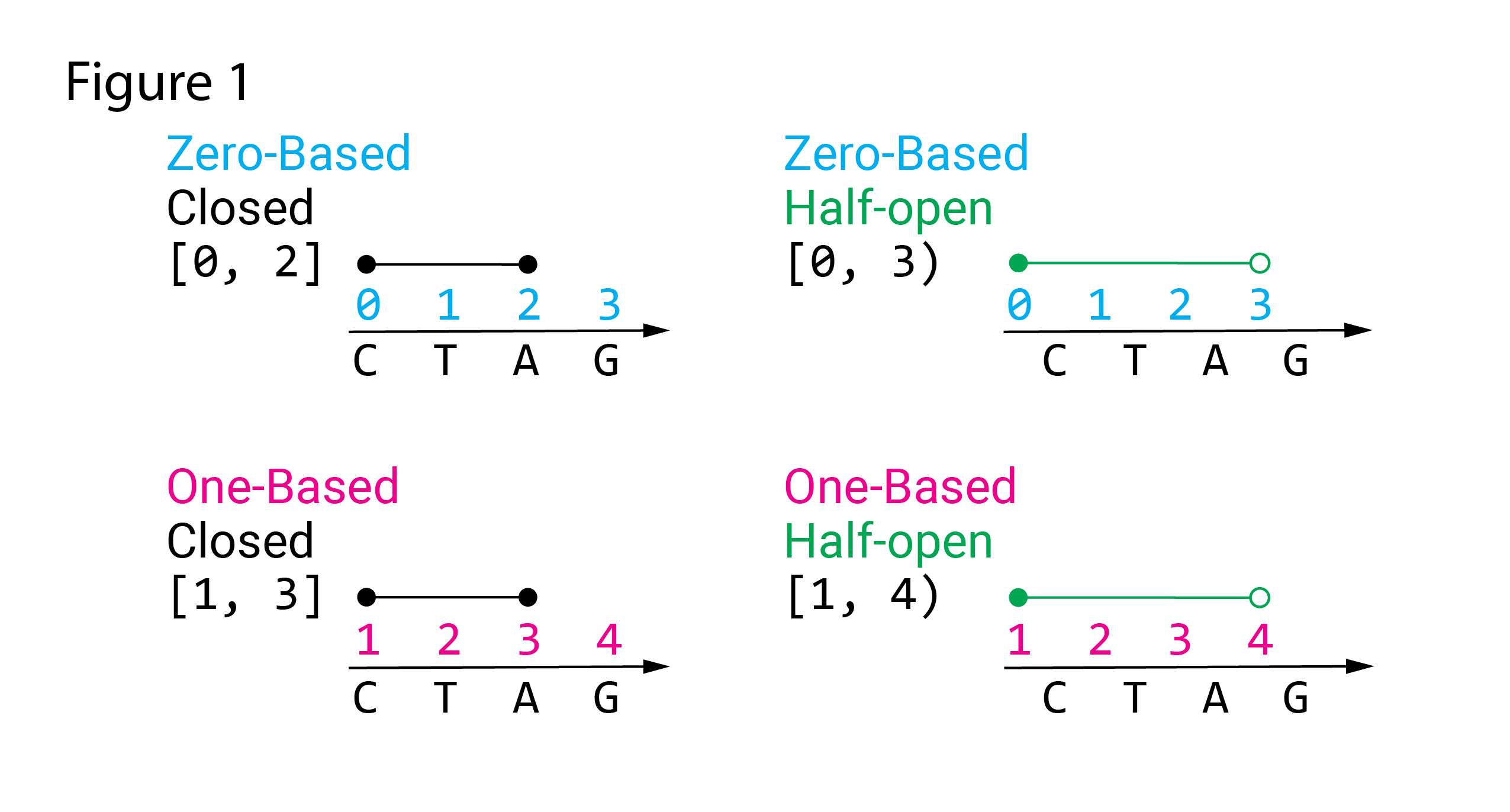}
\caption{Zero- and one-based coordinate basis; closed and half-open intervals. In all four examples, the nucleotide sequence \texttt{CTA} is encompassed by the indicated interval.}
\label{fig:coordinatesystems}
\end{figure}

While  sequence graphs will play an increasingly important role in the future,\cite{Paten01052017} the vast majority of sequence analysis work presently operates on coordinate systems with respect to a linear reference. These coordinate systems are defined on the one-dimensional axis of the reference sequence (which is given an integral or string identifier), with sequence intervals describing a range of positions within the reference. These systems inherently rely on two characteristics: the \textit{basis} (a zero or one indexed counting of the reference sequence), and---when considering intervals---by \textit{openness}: whether the end coordinate of a range is included or excluded (\textit{i.e.}, whether the interval is closed or open). The Cartesian product of the 2 values each of basis and openness yields four distinct potential coordinate systems (Figure \ref{fig:coordinatesystems}).

\begin{figure}[h] 
\includegraphics[width=\columnwidth]{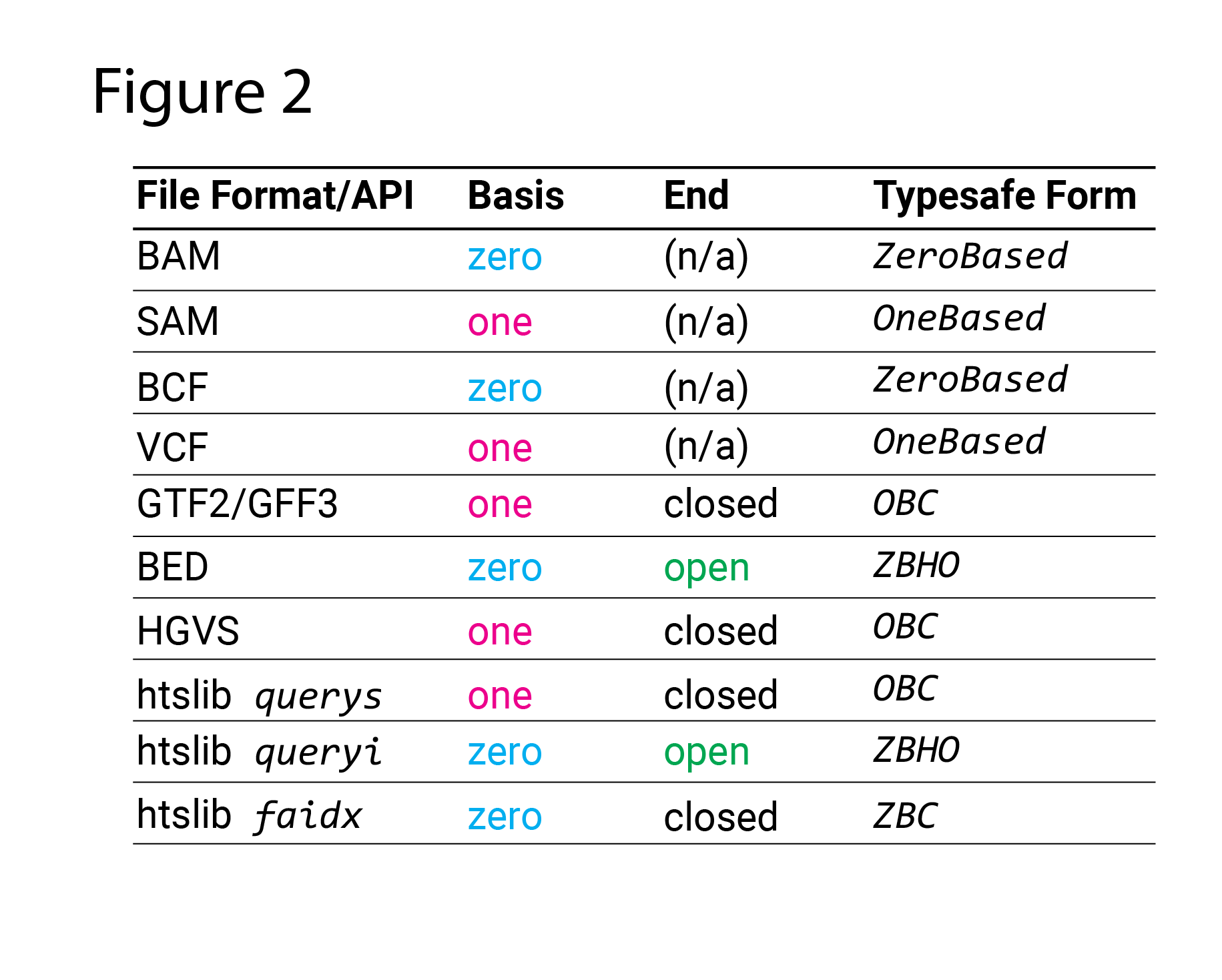}
\caption{File format and API coordinate systems.}
\label{fig:filesandapis}
\end{figure}

The adoption of different systems across different high-throughput sequencing (HTS) file formats may reflect the format's intended consumer, as a way to optimize understanding, programming logic/arithmetic, or optimal computation. For example, many---but not all---text-based formats intended for human consumption use a 1-based system and are closed with respect to intervals, which may feel more natural to the human reader. On the other hand, 0-based systems are most familiar to programmers and are more compatible with default zero-based array indexing in most computer languages. Likewise, while the closed interval $[a,b]$ may read naturally to a human, interval length computations are easier and less prone to error in an open-ended (``half open'') system where the length of an interval $I=[a,b)$ can be computed trivially as $b-a$ without the need for unit adjustment. Figure \ref{fig:filesandapis} summarizes coordinate systems used by common formats.

Unfortunately, the use of alternative systems in different files and application programming interfaces (APIs) which must interoperate sets the stage for subtle off-by-one errors, which are a well-recognized class of problems, but also hard to avoid and sometimes difficult to detect.\cite{CWE193} We propose the use of compilers' type systems to aid bioinformatics software developers and to enforce correctness in a guaranteed way. By applying the principles we describe, the use of a statically typed, compiled programming language for the development of bioinformatics software has the potential to provide these guarantees essentially for free. We provide a reference implementation in D, and proofs of concept in Rust, OCaml, and Python.


\section*{Methods}  

First, we define the possible values of \textit{basis} as ``zero-based'' and ``one-based'', according to whether the first element of the reference sequence is counted from zero or one. Thus, a single coordinate as well as an interval may be ``zero-based'' or ``one-based''.

Second, we define the possible values of \textit{openness} as ``half-open'', in which the end coordinate of the interval is excluded as a member of the range,  or ``closed'', where ``closed'' means that the end coordinate of the interval is encompassed within the the range. Because coordinates represent an integral---not real---valued index into a sequence, the beginning of a genomic interval is always considered to be closed.

Taking these two values on two axes yields four possible combinations, which we will call and abbreviate as follows: zero-based half-open (ZBHO), zero-based closed (ZBC), one-based half-open (OBHO), and one-based closed (OBC). Figure \ref{fig:coordinatesystems} summarizes these four systems.

A basic implementation of coordinate systems might define the types \mbox{\textit{Coordinate}} and \textit{Interval} similar to the following:\\

\begin{lstlisting}[frame=none,basicstyle=\ttfamily\small]
    Coordinate: { basis_0: bool, pos: int64_t }
    Interval:   { closed: bool,
                  start: Coordinate, end: Coordinate }
\end{lstlisting}

Note that enums representing basis and closedness could also be used in place of booleans. Functions that accept (or return) coordinates may then be parameterized as:

\begin{verbatim}
    f(i: Interval, ...) -> j: Interval
\end{verbatim}

Logic within the function should check the basis and openness, and, if necessary, adjust the \texttt{start} and \texttt{end} positions to be concordant with the expected coordinate system. While this provides an improvement over the acceptance of undecorated integral values of unknown provenance, there is not only a runtime cost (which could be substantial if \textit{e.g.} \texttt{f} is called from an inner loop), associated with these checks, but more importantly a tremendous opportunity for error. If at any point in the lifetime of the \texttt{Coordinate} or \texttt{Interval} object a logical error has been made, an invisible off-by-one error may be propagated throughout the program's execution.

Safety could be improved by defining a set of common functions $T_{system}$ where $ system \in \{zbho, zbc, obho, obc\}$ and the $T$ function transforms coordinates to the indicated destination system. However, it is still incumbent upon each function or block of code making use of coordinates and intervals to call these transform functions, their guaranteed usage is not enforced at any level, and they come with the runtime costs of comparison and branching.

\begin{figure}[h]
\includegraphics[width=\columnwidth]{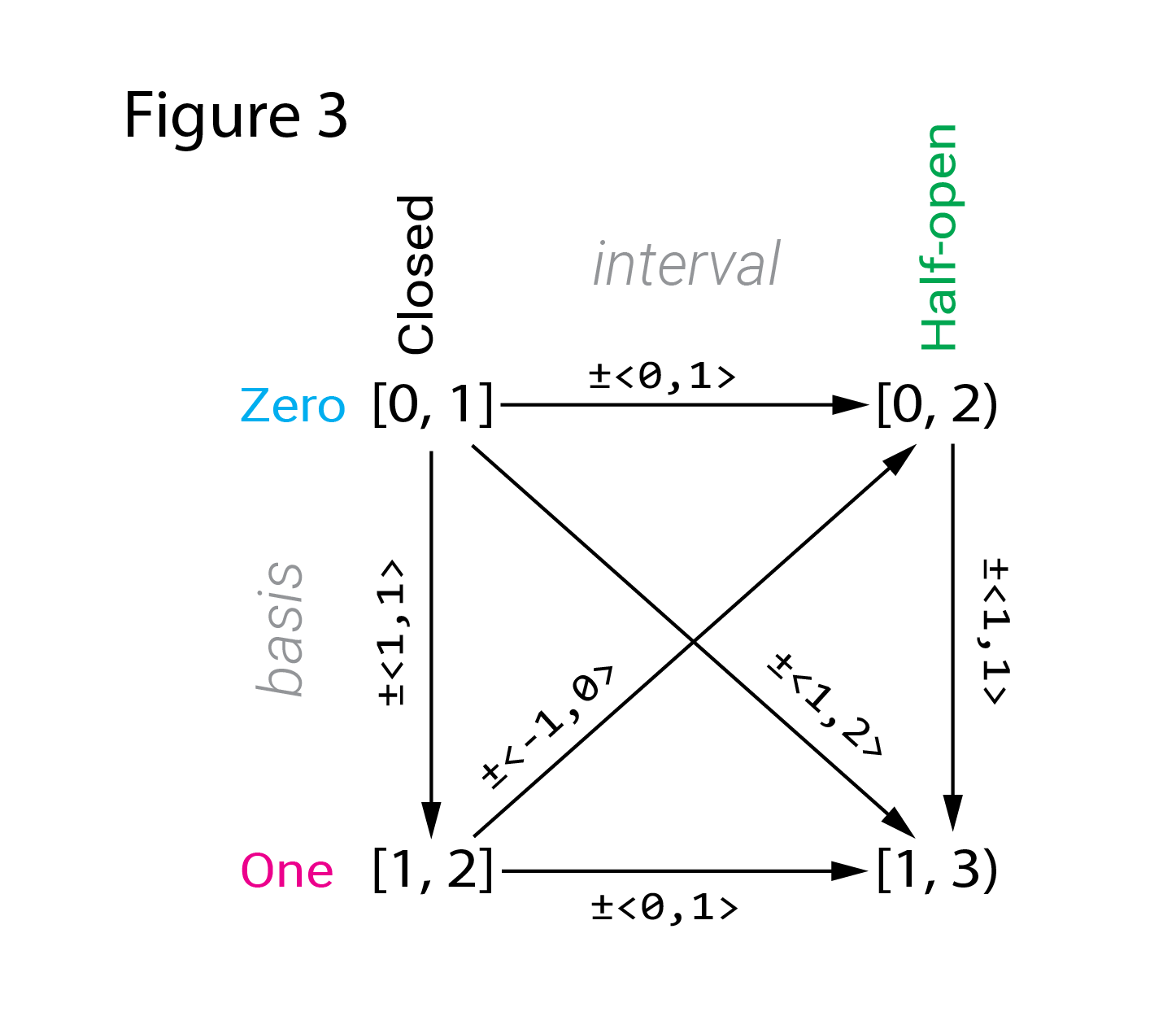}
\caption{Transition graph depicting the vector arithmetic transformations between coordinate systems.}
\label{fig:transitiongraph}
\end{figure}

Instead, consider strongly and statically typed languages with generic and polymorphic features. Define a type class \texttt{Interval} of which there are four types, corresponding to ZBHO, ZBC, OBHO, and OBC coordinate systems, and for which there exists a function, \textit{to}, (or conversely, \textit{from}) that maps between constituent types. Note that the types may be freely interconverted, as there exists a bijective mapping between any two coordinate systems (Figure \ref{fig:transitiongraph}). While these types could all be defined individually, one might instead leverage polymorphic features of the type system, especially in systems where a type can be parameterized on a constant value. 

In C++, D, or similar languages this could be done by templating a single \texttt{Interval} class or struct on some enums defining basis and openness:
\begin{verbatim}
    // C++
    Interval <CoordSystem::zbho> i (0, 100);
    // D
    auto i = Interval!CoordSystem.zbho(0, 100);
    // Rust
    let i = Interval::<ZeroBased, HalfOpen>::from_int(0, 100);
\end{verbatim}

Functions accessing HTS files and records can then be defined to safely take (and return) only the appropriate coordinate/interval type. Passing the wrong coordinate or interval type to a function would thus result in a compile-time error. Alternatively, shifting the burden from library consumer to producer, functions could be defined polymorphically so as to take coordinates or intervals of any type, as long as they include an initial conversion to the correct type within the function body.


We implemented a typesafe coordinate system in the D language using parametric polymorphism on top of our existing \texttt{dhtslib}\cite{gregory_dhtslib_nodate} library, which already possessed an extensive set of HTS file reading and writing interfaces and abstractions. Using D’s extensive template system, contract programming, \texttt{static if} (compile-time branching) capabilities, and other compile-time checks, we defined the type constructors \texttt{Coordinate} and \texttt{Interval}.

\texttt{Coordinate} and \texttt{Interval} are fundamentally structs templated on the values of their basis and openness, and their complete instantiated types are expressed as:

\begin{verbatim}
    Coordinate!Basis.{system}
    Interval!CoordSystem.{system}
\end{verbatim}

(Note that the D type notation \texttt{T!(U, V, \ldots)} is equivalent to \texttt{T<U, V, \ldots>} which is more familiar to users of C++ and Rust.) Type construction is achieved via enums \texttt{Basis}, \texttt{End}, and \texttt{CoordSystem}; strictly speaking, the use of enums means type construction is parameterized on values, rather than types. However, the D compiler assesses template value parameters statically at compile time and the functional result is identical to type construction with types. In Rust, another statically typed compiled language, this is equivalent to const generics, a type system feature that is currently in development.

Values of type \texttt{Coordinate} and \texttt{Interval} can be converted to other types via a polymorphic function \texttt{to} which takes a destination coordinate system type. Based on the source and destination coordinate systems, the underlying integer coordinates are converted statically at compile-time (that is to say, the instructions to increment or decrement are injected) by adding or subtracting depending on the new \texttt{Basis} and \texttt{End} types. Figure \ref{fig:transitiongraph} depicts this interconversion.

Type construction using enum-templated structs yields distinct, concrete types representing different coordinate systems. This API allows for compile-time checking of coordinate system compatibility, and allows for conversion of singular coordinates and coordinate pairs (intervals) from one system to another. The system is considered typesafe as functions may accept coordinates of only one particular system with compilation failure if another type is passed. 

To instead create a function \texttt{fun} that accepts input of all coordinate systems, the function must be polymorphic or generic over the coordinate(s) type. Then in use, \texttt{fun} must be instantiated\footnote{Ideally via monomorphization as in C++, D, or Rust for performance.} with the desired coordinate system type and thereby be parameterized by a known \texttt{Coordinates} type, so type-safety is still achieved. In D this can be inferred, so the function will be automatically instantiated with the correct type. The coordinate system provides compile-time (via the type system) and run-time checks (bounds) to ensure coordinates are valid for their coordinate system; \textit{i.e.}, a one-based system must not contain 0 as a coordinate while a half-open coordinate system’s end coordinate must be greater than its start coordinate.

A complete understanding of D’s compile-time features is not necessary to use the \texttt{dhtslib} Typesafe Coordinate System. To reduce verbosity of type signatures, we provide short type aliases and convenience constructors. For example, to declare integer coordinates as a particular coordinate system, one might declare:

\begin{verbatim}
    auto i = ZBHO(0, 100);  // [0, 100) (zero-based)
    auto j = OBC(1, 100);   // [1, 100] (one-based)
    auto k = i.to!OBC;
    assert(k == j);
\end{verbatim}


\section*{Results}  

We created \texttt{dhtslib},\cite{gregory_dhtslib_nodate} a collection of high-level abstractions, convenience functions, and bindings over the widely-used and low-level \texttt{htslib} C library.\cite{bonfield_htslib_2021} At a high level, \texttt{dhtslib} provides object-oriented readers and writers for BAM/CRAM/SAM, BCF/VCF, BED, and GFF3 with support for region-based queries and filtering via standard indexes.

To effectively deal with the problem of different coordinate systems across these file formats, read and write functions take and receive only typesafe coordinates as described above, which effectively guarantees correctness at compile-time. Through the use of parametric polymorphism and metaprogramming, these readers’ and writers’ function parameters may be coordinate intervals of any type, and are automatically converted --- at compile time --- to the requisite representation for the format in question. Thus, coordinates and ranges from one file acquired via the \texttt{dhtslib} API may be directly and transparently passed to another file, whether or not the two share a common coordinate system, and without any specific intervention on the library user's part.

As a concrete example, consider reading a coordinate range encompassing a gene or exon of interest from a GFF3 gene annotation file. In a typesafe coordinate system-enabled library such as \texttt{dhtslib}, the return value of any GFF3 reader will be of a type encoding its one-based closed (OBC) nature. Suppose then that the software must query a BAM file for alignments falling within the range just queried. In a typesafe coordinate system, the prior result can be passed directly to query function which is generic over the type of input coordinates and at compile time has logic injected to perform any necessary conversion of \textit{basis} and/or \textit{openness}. 

A pseudocode excerpt is shown below.\\

\begin{lstlisting}[language=D,frame=single,basicstyle=\ttfamily\small]
// GFF3 files' coordinates are one-based closed
// regionOfInterest is type Interval!CoordSystem.obc
auto regionOfInterest = gff3file.genebounds("TP53");

// BAM files are zero-based, half-open
// BUT: regionOfInterest from GFF3 can be passed directly
auto reads = bamfile.queryi(regionOfInterest);
\end{lstlisting}


In this way, a developer using \texttt{dhtslib} (and by extension htslib) needn't explicitly be aware of the coordinate space that they receive from one file type or know what coordinate system it should be converted to for use with another file type or API. From \texttt{dhtslib} version 0.12.0 onward, the Typesafe Coordinate System is enforced for all high-level interfaces.

We also explored the feasibility of Typesafe Coordinates in other statically-typed languages, including Rust and OCaml. In Rust, we built Typesafe Coordinates by parameterizing \texttt{Coordinate} and \texttt{Interval} types on either zero-sized structs, or const Enum values (at the time of writing, the latter was only possible in nightly Rust with the ``const generics" feature enabled). While this achieves full compile-time type safety, a runtime type check is still required in generic functions. In OCaml, we likewise parameterized \texttt{Coordinate} and \texttt{Interval} types on zero-sized types and were thus able to avoid runtime type checks, which would have been required if parameterizing on a variant (sum type). OCaml's type inference features led to terse and elegant appearing code, however the lack of function overloading in the context of polymorphism led to more verbose free function names.

Finally, we implemented trivial ``typed" coordinates in Python, an extremely popular dynamically-typed language with object-oriented and functional programming features. Here, we used a class with member variable to track the coordinate system, which permits safe and correct explicit interconversion. However, because function parameters are not typed in Python, there is no runtime safety apart from what class consumers write themselves.

\section*{Discussion}   

Software engineering and bioinformatics are relatively newer disciplines among the applied sciences. In the same way that other engineering fields may apply and benefit from fundamental advances in mathematics and physics, the integrity, safety, and performance of software systems can be improved by applying lessons from computer science. These principles have also made their way to bioinformatics and computational biology, where authors have recognized the Rust language (\url{https://www.rust-lang.org}) for its memory safety attributes and its potential contributions to safer and more correct bioinformatics tools.\cite{koster_rust-bio_2016,perkel_why_2020}  Here, we demonstrate that leveraging the type system and compile-time guarantees of strongly typed languages like D, Rust, and others can reduce the error surface of software in the computational biology and bioinformatics space. These principles could also be extended to more loosely typed language such as Python, albeit with much weaker guarantees and a runtime overhead.

The reference implementation of Typesafe Coordinates as part of \mbox{\textit{dhtslib}\cite{gregory_dhtslib_nodate}} --- to our knowledge the first such polymorphic implementation of this concept in any bioinformatics tooling --- is written in D (\url{https://dlang.org/}), a statically-typed, compiled language. We found D's compile-time introspection and metaprogramming features ideal for construction of such a system. We also implemented Typesafe Coordinates in other languages, including Rust, OCaml, and Python, albeit with language-specific limitations in each of these.

In the thrall of safety, \textit{dhtslib} does not permit circumvention of the system by accepting integer coordinates directly in functions that operate on files or records. This ensures that a software developer cannot create a situation involving off-by-one errors without specific and significant effort. Importantly, this safety is statically guaranteed and therefore comes without any runtime branching penalty. In the case that coordinates must be hardcoded or accepted as untrusted user input, type constructors accept integers, and the explicit initialization of a named type carries distinct advantages. First, it forces the developer in these cases to be sure about the coordinate system (depending on the data’s downstream use). Second, it is implicitly self-documenting. Finally, correctness (or lack thereof) of the coordinates is deterministically transmitted through the call chain and program flow, simplifying debugging.

In conclusion, we have introduced the concept of Typesafe Coordinate Systems for bioinformatics and computational biology, and provided a reference implementation as part of a larger and more comprehensive high-throughput sequencing package \texttt{dhtslib}. We have also provided additional proofs-of-concept in various languages including Rust, an emerging language for bioinformatics. We hope that these safety concepts will be adopted by others and implemented either as standalone packages (leveraging macros and metaprogramming), or directly integrated into popular libraries and packages such as SeqAn,\cite{reinert_seqan_2017} BioPython,\cite{cock_biopython_2009}, Rust-Bio,\cite{koster_rust-bio_2016} and hts-nim.\cite{pedersen_hts-nim_2018} Using type systems to eliminate error in specific problem domains is a powerful safety concept and should become routine in bioinformatics.

\section*{Acknowledgements} 

The authors thank Brent Pedersen and Nicholas Wilson for critically reading the draft manuscript, and Seb Mondet for reviewing the OCaml code.

\newpage
\printbibliography

\newpage
\section*{Figures}

\end{document}